\RequirePackage{fix-cm}
\documentclass[smallextended]{svjour3}       
\smartqed  
\usepackage[utf8]{inputenc}
\usepackage{amsmath}

\usepackage{braket}
\usepackage{pgfplots}
\usepackage{csquotes}
\usepackage{float}
\usepackage{multirow}
\usepackage{dcolumn}
\usepackage[colorlinks=true, citecolor=blue]{hyperref}
\usepackage{hhline}
\usepackage{amssymb}
\usepackage{amsmath}
\usepackage{graphicx}
\usepackage{subfigure}

\usepackage{braket}
\usepackage{graphicx}
\usepackage{float}
\usepackage{pgfplots}
\usepackage{dsfont}
\usepackage{subfloat}
\usepackage{csquotes}
\usepackage{multirow}
\usepackage{dcolumn}
\usepackage[colorlinks=true,citecolor=blue]{hyperref}
\usepackage{tikzfig}
\usepackage{tikzstyles}
\usepackage{circuitikz}
\usepackage{braket}
\usepackage{bbold}
\begin{document}

\title{Demonstration of the No-Hiding Theorem on the 5 Qubit IBM Quantum Computer in a Category Theoretic Framework}

\titlerunning{Demonstration of the No-Hiding Theorem...}

\author{Amolak Ratan Kalra \and Navya Gupta \and Bikash K. Behera \and Shiroman Prakash  \and Prasanta K. Panigrahi}

\institute{Amolak Ratan Kalra \at Department of Electrical Engineering, Dayalbagh Educational Institute, Agra 282005, Uttar Pradesh, India\\ \email{amolakratankalra@gmail.com} \and Navya Gupta  \at Department of Physics, Indian Institute of Technology, Kanpur 208016, Uttar Pradesh, India\\ \email{navyag@iitk.ac.in} \and Shiroman Prakash \at   Department of Physics and Computer Sciences, Dayalbagh Educational Institute, Agra 282005, Uttar Pradesh, India \\ \email{shiroman@gmail.com} \and Bikash K.Behera \at Department of Physical Sciences, Indian Institute of Science Education and Research Kolkata, Mohanpur 741246, West Bengal, India \and Prasanta K. Panigrahi \at Department of Physical Sciences, Indian Institute of Science Education and Research Kolkata, Mohanpur 741246, West Bengal, India\\ \email{pprasanta@iiserkol.ac.in}}

\date{Received: date / Accepted: date}

\maketitle

\begin{abstract}
The quantum no-hiding theorem, first proposed by Braunstein and Pati [Phys. Rev. Lett. \textbf{98}, 080502 (2007)], was verified experimentally by Samal \emph{et al.} [Phys. Rev. Lett. \textbf{186}, 080401 (2011)] using NMR quantum processor. Till then, this fundamental test has not been explored in any other experimental architecture. Here, we demonstrate the above no-hiding theorem using the IBM 5Q quantum processor. Categorical algebra developed by Coecke and Duncan [New J. Phys. \textbf{13}, 043016 (2011)] has been used for better visualization of the no-hiding theorem by analyzing the quantum circuit using the ZX calculus. The experimental results confirm the recovery of missing information by the application of local unitary operations on the ancillary qubits. 
\end{abstract}

\keywords{IBM Quantum Experience, No-Hiding Theorem, Quantum Information}

\section{Introduction}
IBM has developed 5-qubit and 16-qubit superconducting qubit-based quantum computers (ibmqx2, ibmqx4, ibmqx5) which have been released to the research community at large via a web-based interface called \textit{IBM Quantum Experience} \cite{qnoh_IBM} (IBM QE). It is world's first commercial quantum computing service provided by IBM and permits a user to run quantum algorithms via the IBM cloud and to implement quantum circuits. Using this web interface researchers have run a variety of quantum computing and quantum information experiments and demonstrations. These include experiments/demonstrations in the field of quantum information \cite{qnoh_LeggetIBM,qnoh_MerminIBM}, condensed matter physics \cite{qnoh_cond1,qnoh_cond2}, quantum artificial intelligence \cite{qnoh_ai1}, quantum gravity \cite{qnoh_qg1}, quantum simulation \cite{npjQIViyuela2018,qnoh_qg2}, quantum cryptography \cite{qnoh_BeheraQIP2017,qnoh_RGBiswas2018}, quantum error correction \cite{qnoh_WoottonIBM,qnoh_Bikash3,qnoh_VuillotQIC2018,qnoh_arXivHarper2018}, quantum entanglement based protocols \cite{qnoh_SisodiaQIP2017,qnoh_Anirban,qnoh_Abhishek,qnoh_EntropicIBM,qnoh_Bikash2} and quantum cloud computing \cite{qnoh_CloudIBM,qnoh_LinkeIBM} to name a few.  

Two well-known and important qualitative features of quantum information are that, unlike classical information, it cannot be cloned \cite{qnoh_no-cloning} or deleted \cite{qnoh_no-deleting}. A closely-related, and equally important feature of quantum information, proven by Braunstein and Pati \cite{qnoh_no-hiding} is that quantum information also cannot be ``hidden". This is the so called ``quantum no-hiding theorem", which is relevant to questions about thermalization in quantum systems \cite{qnoh_thermalization}, the black hole information loss paradox \cite{qnoh_HwakingNature1974} and other areas in quantum information where state randomization plays a role, e.g. private quantum channels \cite{qnoh_private}. Quantum Information is fragile and any interaction or disturbance to the system may lead to loss of information. The no-hiding theorem addresses this issue of information loss and says that no information can be hidden in correlations between a pair of systems. If any information is missing from a physical system, it must have moved to somewhere else and cannot be hidden in correlations between the physical system and the environment. There are two versions of the no-hiding theorem, one which deals with perfect hiding processes and another in which the hiding process has imperfections. The imperfection arises mainly due to imperfections in the encoding process. 

In the present work, we discuss about both processes and provide a complete experimental verification of the no-hiding theorem with a perfect hiding process. We propose a quantum circuit which can be experimentally implemented to investigate processes with imperfect hiding processes. As for the experimental architecture, we use IBM Quantum Experience platform to test and demonstrate the quantum no-hiding theorem. We remark that this work is inspired by the first experimental test of no-hiding theorem using an NMR based quantum computer by Samal \emph{et al.} \cite{qnoh_no-hiding-NMR}. 

\section{Review of The Quantum No-Hiding Theorem}
\label{qnoh_no-hiding-review-section}

In classical physics, it is possible to ``hide" information in correlations. A simple example is the one-time pad \cite{qnoh_one-time-pad}, used in cryptography. Consider a message which is a binary string: $M=M_1M_2M_3 \ldots$, and a secret key which is also a binary string of the same length: $K=K_1K_2K_3 \ldots$. Create an encoded message $M'$ by performing a bitwise XOR of $K$ on $M$ (i.e., flip the $M_i$ if $K_i=1$, otherwise leave $M_i$ unchanged). Shannon proved that an agent with access to only the encoded string $M'$, and not the key $K$, has no information about the original message $M$ \cite{qnoh_shannon1949}. Of course, an agent with access to only the key $K$, and not the encoded string $M'$, also has no information about the original message $M$. To obtain any information about the original message $M$, an agent requires at least some information about the correlations between the key $K$ and the encoded string $M'$. In this sense, one can say that the information originally contained in $M$ has been hidden in correlations between the encoded string $M'$ and the key $K$. Braunstein and Pati proved that, in quantum mechanics, hiding information in such a way is impossible \cite{qnoh_no-hiding}. 

Specifically, let us define a \textit{bleaching} process as a process which transforms the state of a quantum system $M$ to the maximally mixed state (or, more generally, any fixed density matrix $\rho$), regardless of what the initial state of the system was. After bleaching, an agent with access to only the  system $M$ has no information about its original state. Bleaching processes are clearly not unitary, but can take place if $M$ is an open quantum system coupled to an environment. After bleaching, the quantum information previously contained in $M$ cannot be truly lost (assuming the universe is a closed quantum system), so it is natural to ask where in the universe it resides. The no-hiding theorem states that, in any bleaching process, the quantum information must be transferred completely to the environment, and (unlike the example of the classical one-time pad) cannot reside within correlations between the environment and the original system $M$. In particular, an agent who has access to only the environment and not the bleached system $M$, can, in principle, always completely recover the quantum information initially stored in $M$, via an appropriate unitary transformation. For a large environment, this unitary transformation may be hopelessly complicated. However, for a smaller ``environment" consisting of a few qubits, it is possible to explicitly construct the required unitary transformation and thereby directly demonstrate the no-hiding theorem, as we show in the following sections.

\section{Circuit to Demonstrate the No-Hiding Theorem}

The quantum circuits demonstrating the no-hiding theorem are depicted in Fig. \ref{qnoh_Fig1(a)} and \ref{qnoh_Fig1(b)}. Fig. \ref{qnoh_Fig1(a)} illustrates the circuit for bleaching the system $M$, which is represented by the single qubit state $|\psi\rangle$. In order to bleach one qubit, we require an ``environment" consisting of at least two ancilla qubits. A 3-qubit unitary ``randomization" operator $\mathbf{U}$ that achieves bleaching is the controlled operation that applies one of the four Pauli operators  $\mathds{1}$ (the identity), $\mathbf{X}$, $\mathbf{Y}$ or $\mathbf{Z}$ to $M$, depending on whether the two ancilla qubits are in the state $\ket{00}$, $\ket{01}$, $\ket{10}$ or $\ket{11}$ respectively. This can be written as

\begin{equation}\label{qnoh_Eq1}
    \mathbf{U} =  \mathds{1} \otimes \ket{00}\bra{00} + \mathbf{X} \otimes \ket{01}\bra{01} + i \mathbf{Y}\otimes \ket{10}\bra{10}+\mathbf{Z}\otimes \ket{11}\bra{11}
\end{equation}

\begin{subfigures}
\begin{figure}[]
\includegraphics[scale=2.0]{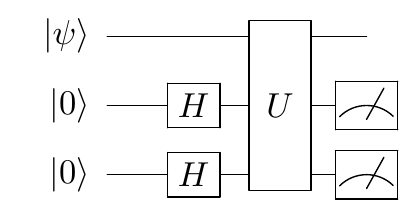}
\caption{\textbf{Erasure circuit.} The two ancillas are prepared in the $|+\rangle$ state by applying two Hadamard gates. Then the unitary operator $U$ is applied on the three qubit state, $|\psi\rangle|00\rangle$ for randomizing $|\psi\rangle$. If the two ancillas are discarded, the first qubit is left in a maximally mixed state. It is to be noted that the unitary operation shown in the figure has been chosen as given in Eq. \eqref{qnoh_Eq2}}
\label{qnoh_Fig1(a)}
\end{figure}

\begin{figure}
\includegraphics[scale=2.0]{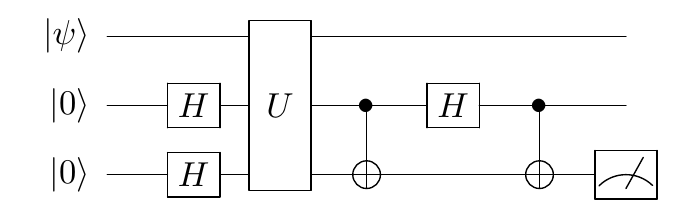}
\caption{\textbf{Quantum circuit demonstrating no-hiding theorem.} In this circuit, the erasure circuit shown in Fig. \ref{qnoh_Fig1(a)} is followed by a decoding circuit, which recovers the state $|\psi\rangle$ from the two ancilla qubits. Essentially, the decoding operation transfers the state $|\psi\rangle$ to the third qubit. It is to be noted that the unitary operation shown in the figure has been chosen as given in Eq. \eqref{qnoh_Eq2}}.
\label{qnoh_Fig1(b)}
\end{figure}

\end{subfigures}

Suppose we initialize the two ancilla qubits in the state $\frac{1}{2} \left(\ket{00}+\ket{01}+\ket{10}+\ket{11}\right)$. Then after applying the unitary randomization operator $\mathbf{U}$, and tracing out the ancilla qubits, it is observed that the state of $M$ is the maximally mixed state, regardless of its initial state $\ket{\psi}$. Hence, we have bleached $M$. The no-hiding theorem states that we can recover $\ket{\psi}$, acting exclusively on the two ancillas. As mentioned earlier, for a larger environment, and a complicated randomization operator $\mathbf{U}$, this might be an intractable task, however for our relatively small system, it is easier to determine the two-qubit unitary operator that decodes the apparently lost information from the environment. We achieve this using the pictorial ZX calculus of Coecke and Duncan \cite{qnoh_Bob1} given in Section \ref{qnoh_ZX-section}. The information recovery circuit is shown in Fig. \ref{qnoh_Fig1(b)}.
 
\section{IBM QE Implementation}
    
\begin{figure}[H]
\includegraphics[scale=.4]{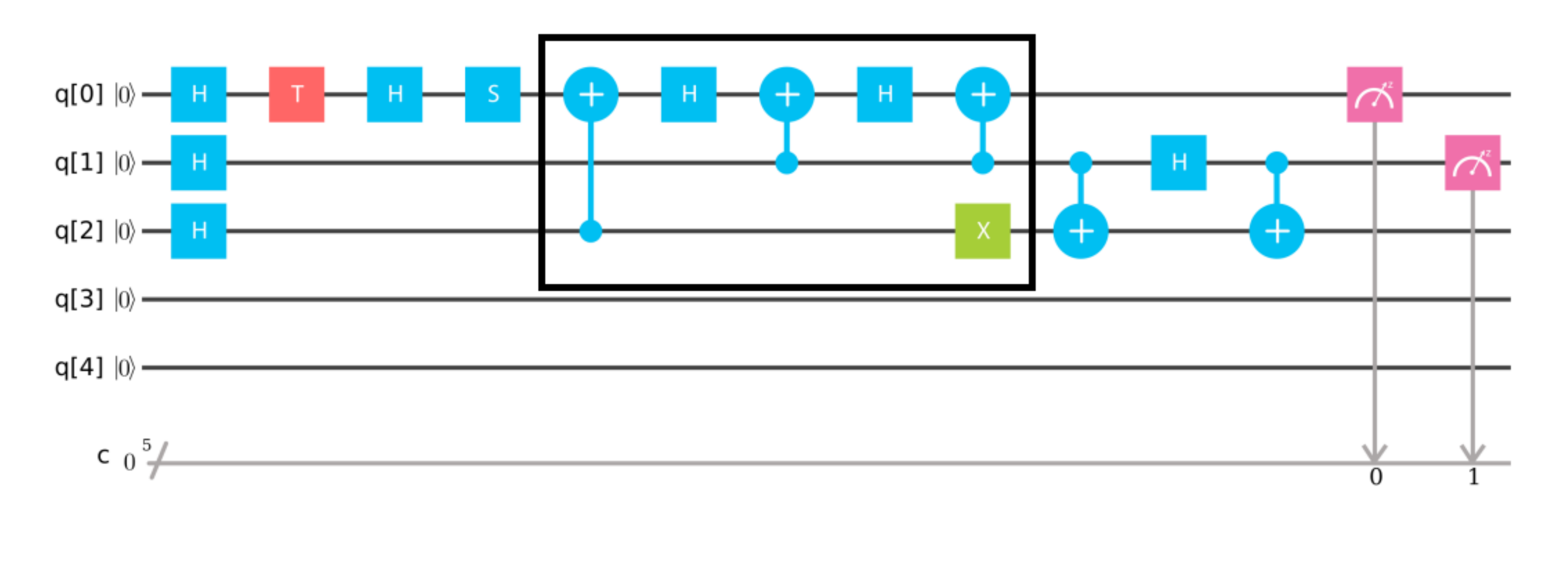}
\caption{\textbf{Quantum circuit illustrating the no-hiding theorem.} The black box represents the quantum operation corresponding to the unitary operator $\mathbf{U}$ given in Eq. \eqref{qnoh_Eq2}. The circuit shown works correctly for general states also.}
\label{qnoh_Fig2}
\end{figure}

Let us first prepare the initial state of the qubit $q[0]$ in $\ket{\psi}$ state, which is taken as $\ket{\psi}=cos(\pi/8)|0\rangle+sin(\pi/8)|1\rangle$. The above state is prepared with the sequential operation of H, T, H and S gates, where, H is the Hadamard operation, T and S gates are phase gates whose matrices are; T = [1, 0;0, e$^{i\pi/4}$] and S = [1, 0;0, i]. Then according to the quantum circuit (Fig. \ref{qnoh_Fig1(a)}), each ancilla qubits (q[1] and q[2]) are prepared in the equal superposition state $|+\rangle$ by applying Hadamard operations on them. The randomization unitary operation is then designed by using a sequence of controlled-NOT, Hadamard and X gates as shown in Fig. \ref{qnoh_Fig2}. It can be mentioned that the circuit shown in the Fig. \ref{qnoh_Fig2} works for any arbitrary state, a$|0\rangle$+b$|1\rangle$, where $|a|^2+|b|^2=1$. Here, we use the following unitary operator for randomization: 

\begin{equation}\label{qnoh_Eq2}
    \mathbf{U} =  \mathds{1} \otimes \ket{01}\bra{00} + \mathbf{X} \otimes \ket{00}\bra{01} - i \mathbf{Y}\otimes \ket{11}\bra{10}-\mathbf{Z}\otimes \ket{10}\bra{11}
\end{equation} 

It can be mentioned that the unitary operator designed here is not the same as the randomization operator given in Eq. \eqref{qnoh_Eq1}. However, it performs the same task to demonstrate the no-hiding theorem, i.e., after the application of randomization operator, the first two qubits 
q[0] and q[1] are entangled in one of the Bell states and the initial arbitrary state $\ket{\psi}$ is transferred to the third qubit q[2]. The brief calculation is given as; $CNOT_{12}H_{1}CNOT_{12}\mathbf{U}|\psi\rangle|+\rangle|+\rangle$ $\rightarrow \frac{|01\rangle+|10\rangle}{\sqrt{2}}|\psi\rangle$, where $CNOT_{ij}$ denotes the operation of controlled-Not gate from q[i]$\rightarrow$q[j], where q[i] is the controlled qubit and q[j] is the target qubit. Similarly, $H_{i}$ denotes the application of Hadamard operation on the q[i] qubit. It is to be noted that there are many different randomization operators which one may use, provided the above condition is satisfied. 

\begin{figure}[]
\includegraphics[scale=0.6]{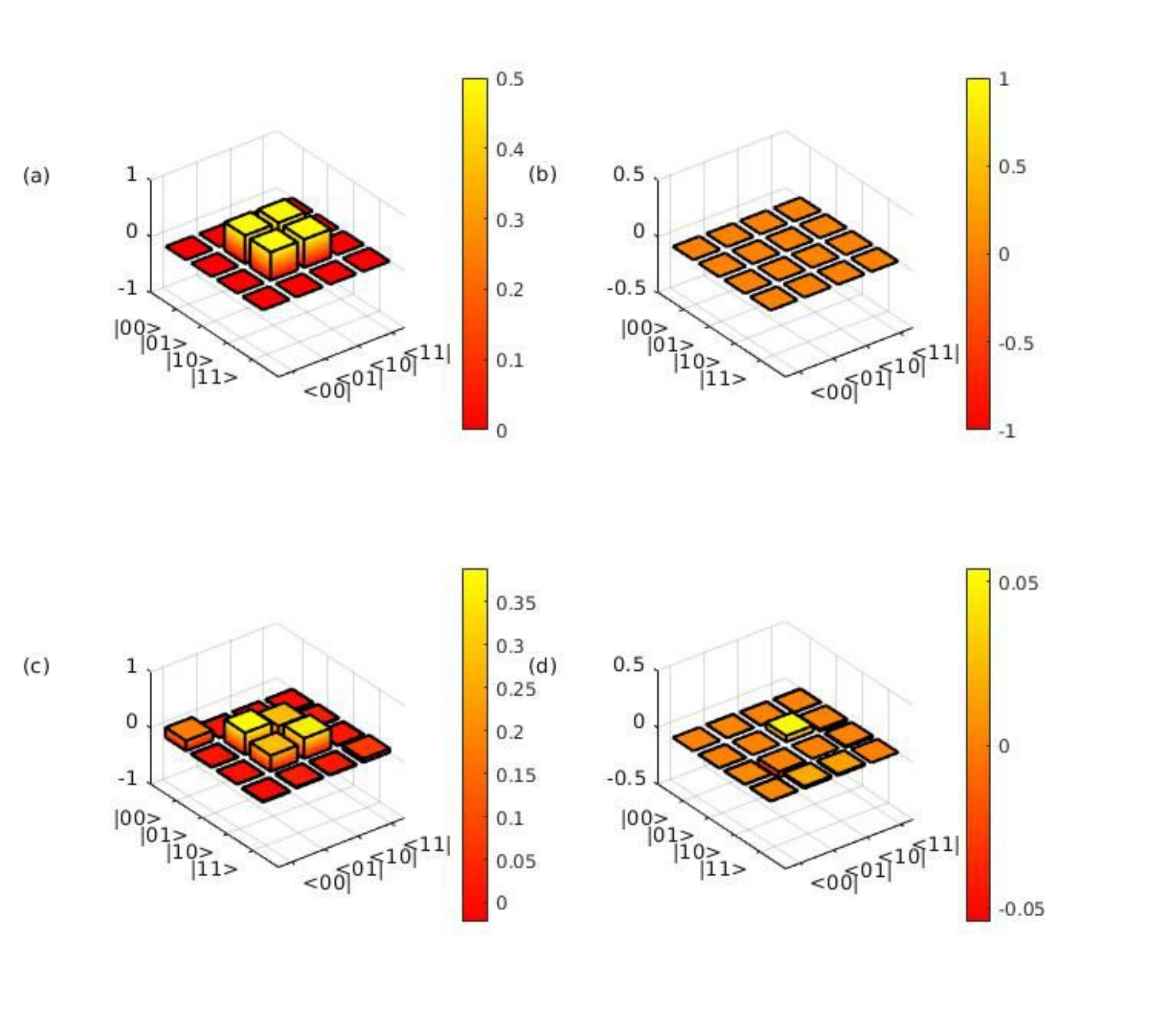}
\caption{\textbf{Graphs showing the tomography of state in qubits q[0] and q[1] ( See Fig. \ref{qnoh_Fig2}).} Depicting real (left) and imaginary (right) parts of the reconstructed theoretical (a,b) and experimental (c,d) density matrices. The fidelity is found to be $0.9905$}
\label{qnoh_Fig3}
\end{figure}

\begin{figure}[]
\includegraphics[scale=0.5]{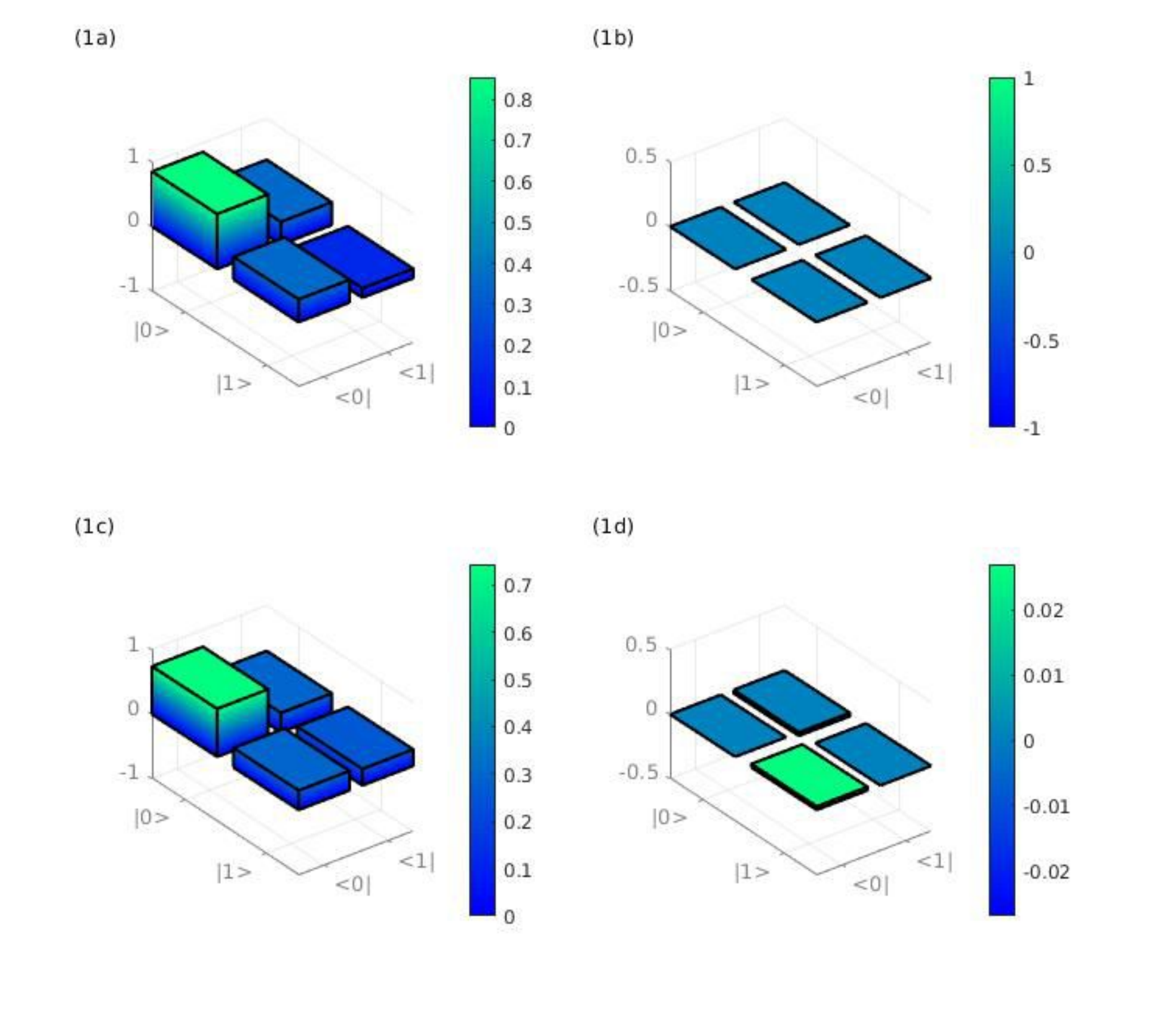}
\caption{\textbf{Graphs illustrating the tomography of state in qubit q[2] (See Fig. \ref{qnoh_Fig2}).} Depicting real (left) and imaginary (right) parts of the reconstructed theoretical (1a,1b) and experimental (1c,1d) density matrices. The fidelity is found to be $0.9967$.}
\label{qnoh_Fig4}
\end{figure}

The quantum circuit (Fig. \ref{qnoh_Fig2}) is designed on the quantum chip, `ibmqx4' with optimized version and the experimental results are obtained with 8192 shots. We perform quantum state tomography by measuring the qubits in different bases and plot the density matrices. Fig. \ref{qnoh_Fig3} shows the theoretical (a,b) and experimental (c,d) density matrices of the quantum state stored on the q[0] and q[1] qubits ($\frac{|01\rangle+|10\rangle}{\sqrt{2}}$). The fidelity for the above quantum state is calculated to be $0.9905$. As the quantum state, $|\psi\rangle$ is transferred to the third qubit q[2], by measuring the third qubit and performing state tomography (Fig. \ref{qnoh_Fig4}), the fidelity was found to be $0.9967$.

\section{Imperfect Hiding}

The bleaching process demonstrated in the previous sections can be represented by the map $\varepsilon_1(\rho)=\frac{I}{2}=\rho+ \frac{1}{4}(X \rho X + Y \rho Y + Z \rho Z)$, acting on the input state $\rho$. We could also consider a more general process represented by the map $\varepsilon^p_2\big(\rho\big)=p\frac{I}{2}+(1-p)\rho=(1-\frac{3p}{4})\rho+\frac{p}{4}(X \rho X + Y \rho Y + Z \rho Z)$, where $0\leq p \leq 1$. For $p=1$, we recover the perfect bleaching process $\varepsilon^1_2(\rho)=\varepsilon_(\rho)$, while for $p=1-\epsilon$, where $0 < \epsilon \leq 1$, we obtain an imperfect bleaching process. Braunstein and Pati \cite{qnoh_no-hiding} have shown that the hiding process is robust to such imperfections. To make this mathematically precise, we note that for this imperfect bleaching process, the trace distance between the imperfectly hidden state and the perfectly hidden state is given by:

\begin{equation}\label{qnoh_Eq3}
    \frac{1}{2}tr|\varepsilon^p_2(\rho)-\varepsilon_1(\rho)|=\frac{1-p}{2}tr|\rho-\frac{I}{2}|=\frac{\epsilon}{2}tr|\rho-\frac{I}{2}|=\frac{\epsilon}{2}|\lambda_1-\lambda_2|\leq \frac{\epsilon}{2}
\end{equation}

where $\lambda_1$ and $\lambda_2$ are the eigenvalues of $\rho$. When $\rho$ is a pure state, the trace distance is equal to $\epsilon/2$. If $F(\sigma,\rho)$ represents the fidelity between the states $\sigma$ and $\rho$, it follows that:

\begin{equation}\label{qnoh_Eq4}
    F(\varepsilon_2^p(\rho),\varepsilon_1(\rho))=\frac{1}{\sqrt{2}}tr \sqrt{\varepsilon_2^p(\rho)} \geq  1 - \frac{\epsilon}{2}
\end{equation}

Eq. \eqref{qnoh_Eq4} can be used to establish that the final composite states of the system and the ancilla in these two cases, given by $|\psi_{perfect}\rangle$ and $|\psi_{imperfect}\rangle$, overlap strongly i.e,:

\begin{equation}\label{qnoh_Eq5}
    \langle \psi_{perfect} | \psi_{imperfect} \rangle \geq 1 - \frac{\epsilon}{2}
\end{equation}

This is a statement of the robustness of the hiding process to imperfections. In the following experiment, instead of measuring the global state, we demonstrate the robustness to perturbations by measuring the final state $\varepsilon_2^p(\rho)$ of the system for various values of $p$, and calculating their fidelity and trace distance with respect to $\epsilon_1(\rho)=\frac{I}{2}$.\\

\begin{figure}
\includegraphics[scale=.15]{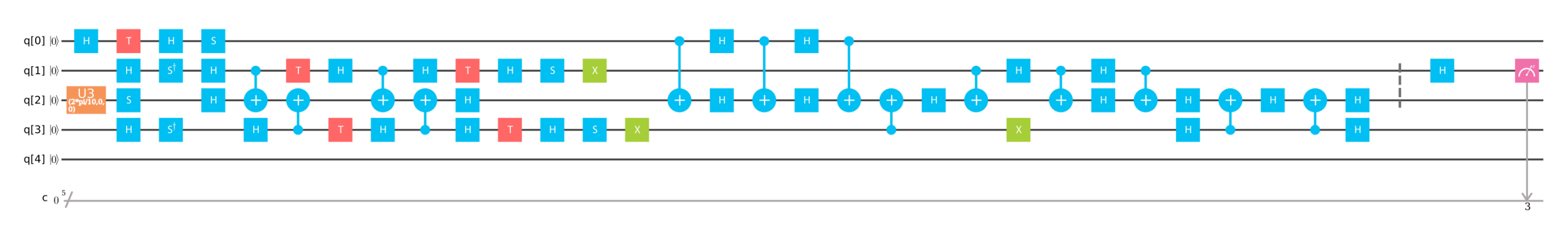}
\caption{\textbf{IBM quantum circuit to demonstrate imperfect bleaching.} The two unitary operators act as the bleaching operators, this circuit does the decoding operation also.}
\label{qnoh_imperfect}
\end{figure}

This time the experiment is performed on ibmqx2 with 1024 shots for each measurement. The IBM quantum circuit which implements the map $\varepsilon_2^p$ as well as the decoding is shown in Fig. \ref{qnoh_imperfect}. The only thing which is different in this case as compared to the previous one is the initial state of the ancilla qubits. This time the ancilla qubits $q[1]$ and $q[3]$ are prepared in the state $\sqrt{1-\frac{3p}{4}}\ket{00}+\sqrt{\frac{p}{4}}(\ket{01}+\ket{10}+\ket{11})$
by applying two controlled-Hadamard gates to the initial state $\ket{00}$, the control being the additional qubit $q[2]$ which is prepared in the state $\sqrt{1-p}\ket{0}+\sqrt{p}\ket{1}$ using the $U3$ gate. The state of the input qubit $q[0]$ is taken as $\ket{\psi}=cos(\pi/8)|0\rangle+sin(\pi/8)|1\rangle$  and is prepared using the sequence of $H$,$T$, $H$, and $S$ gates, as in the previous experiment. Note that due to restrictions on placement of CNOT gates in the `ibmqx2' architecture, we had to implement some qubit swap operations in the middle of the circuit. The two ancilla qubits are finally transferred to wires 3 and 4 ($q[2]$ and $q[3]$), while the system qubit is transferred to wire 2 ($q[1]$). Finally, we perform quantum state tomography of the qubit on wire 2. Fig. \ref{qnoh_imperfect} illustrates the measurement of $q[1]$ in the $X$ basis for $p=sin^2(\pi/10)$.\\

\begin{figure}
\includegraphics[scale=.2]{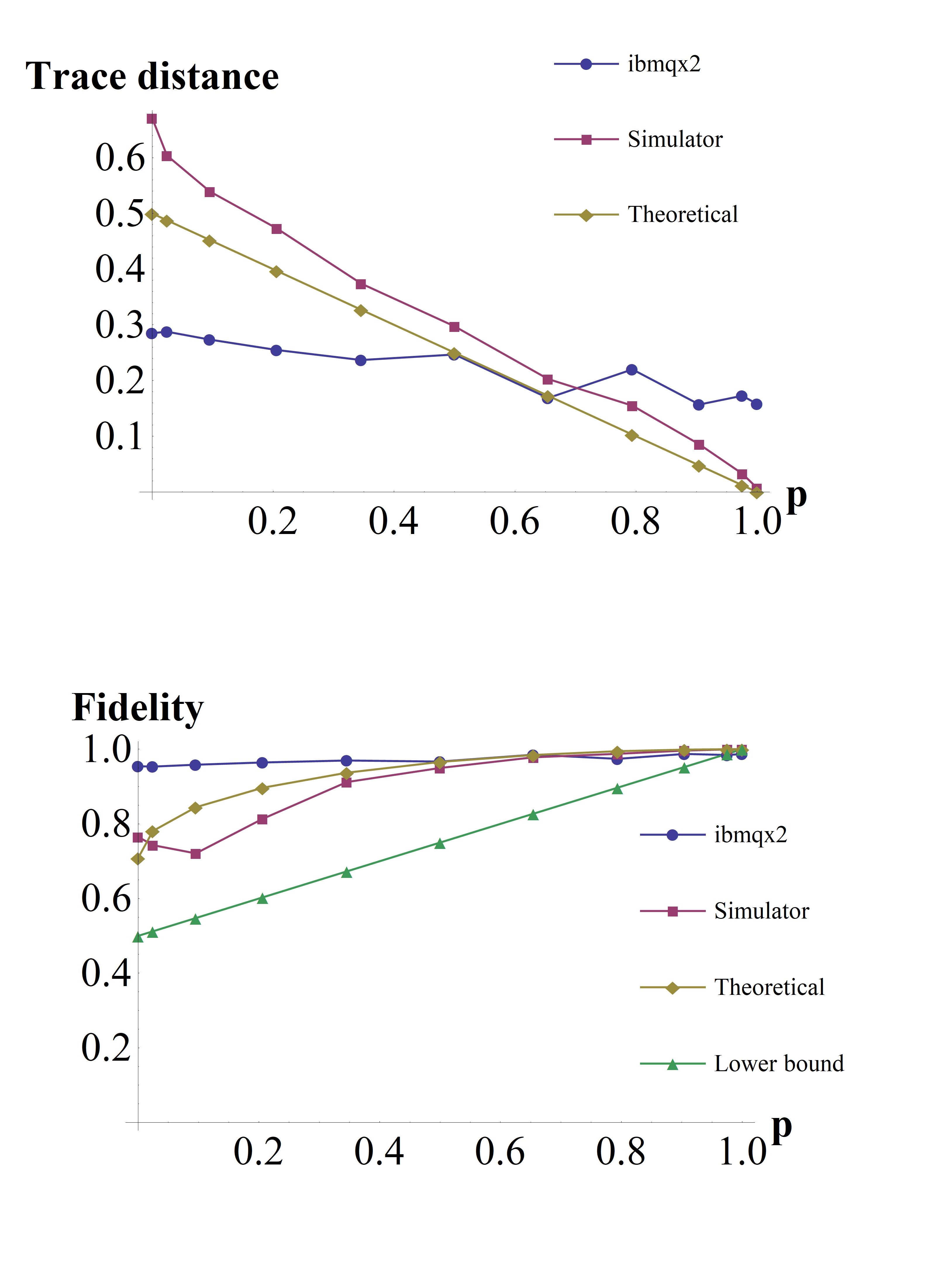}
\caption{\textbf{Trace distance and fidelity between $\varepsilon_1(\rho)$ and $\varepsilon_2^p(\rho)$ as a function of $p$}}
\label{qnoh_Fig5a}
\end{figure}

The results of the experiment are shown in Fig. \ref{qnoh_Fig5a}. As also shown in Eq. \eqref{qnoh_Eq3}, the theoretical trace distance is linearly proportional to $p$ and falls from $0.5$ to $0$ as $p$ increases from $0$ to $1$. The trace distance corresponding to the simulated state is higher than the theoretical trace distance, but it exhibits the linear trend fairly well. It is important to note that the states obtained through tomography using the simulator were nonphysical (not positive definite) for smallest values of $p$ ($p=0.095491503,0.024471742,0$). This is reasonable because when $p$ tends to zero, the final state of the system becomes purer, and the probability of measuring a nonphysical density matrix rises. In the top graph, this fact is manifested by the value of trace distance being greater than $0.5$ for smaller values of $p$. When the experiment is performed on ibmqx2, the final state does not show as strong a dependence on the parameter $p$. This is perhaps because the effect of the deliberate imperfection introduced is washed out by experimental noise. In the bottom graph, in addition to plotting the fidelities obtained from ibmqx2, the simulator, and theoretical calculations, we plot the lower bound for fidelity given by Eq. \eqref{qnoh_Eq3}.   

\section{Diagrammatic Derivation}
\label{qnoh_ZX-section}

Here, we analyze a randomization circuit presented above using the $ZX$-calculus \cite{qnoh_Bob1}. Please note that this randomization circuit is slightly different from the one above but does the same operation. $ZX$-calculus can be understood either as a convenient set of pictorial rules for demonstrating the equivalence of certain quantum circuits, or as an alternative axiomatization of quantum mechanics using the framework of dagger symmetric monoidal categories.\\

\begin{figure}
\centering
\includegraphics[width=0.8\linewidth]{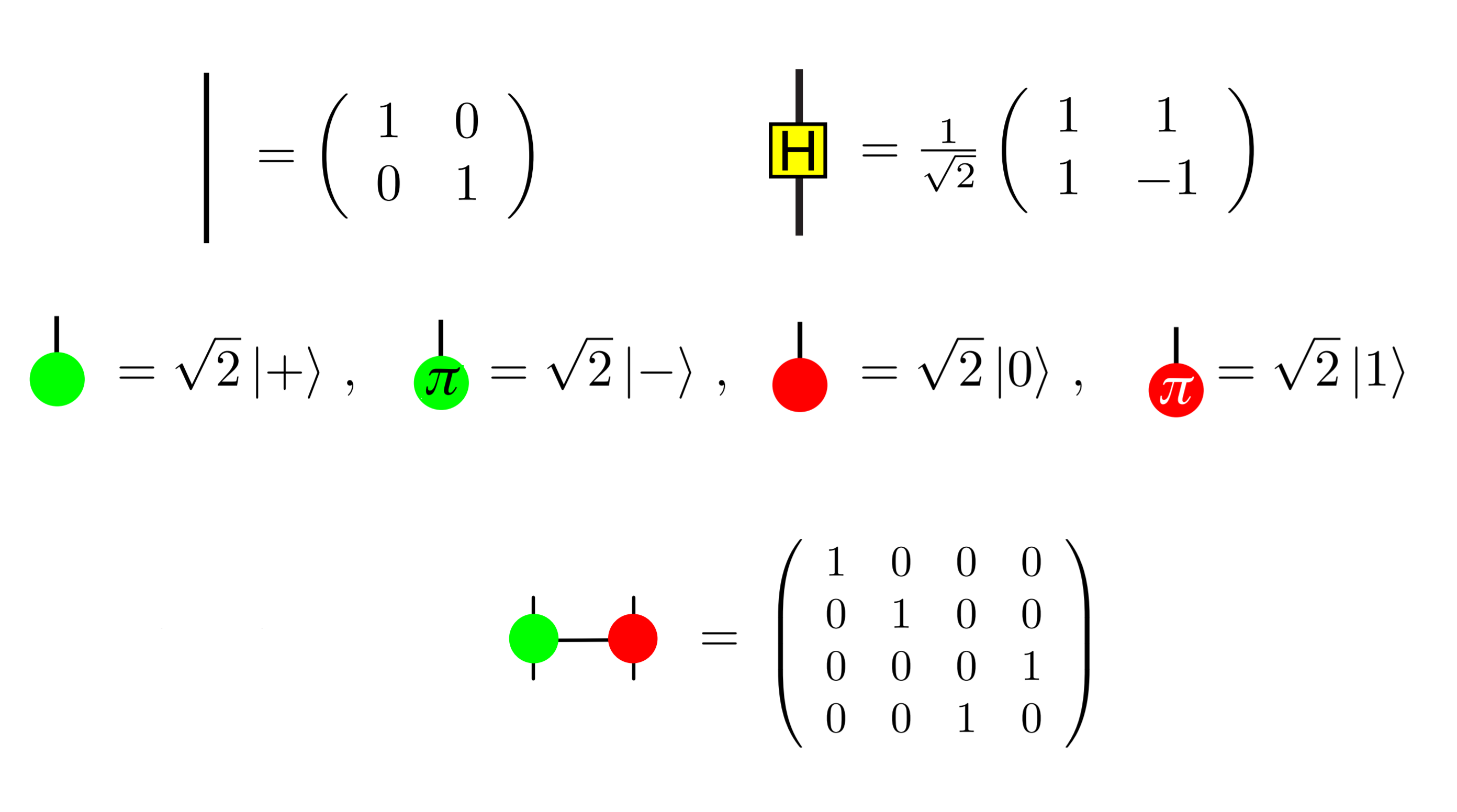}
\caption{Relevant components and their matrix representations.}
\label{qnoh_Fig7}
\end{figure}

The basic elements of the $ZX$-calculus are red and green spiders (represented by red and green dots with $n$ incoming wires and $m$ outgoing wires), and the Hadamard gate (represented by a yellow square, with $H$ written in it). Green spiders act as copy/delete operators in the $Z$ basis (i.e., the computational basis $\ket{0}$, $\ket{1}$) and red spiders act as copy/delete operators in the $X$ basis (i.e., $\ket{+}$, $\ket{-}$).  So for example, a red dot with 2 incoming wires and 1 outgoing wire represents the operator $\ket{+}\bra{+}\bra{+}+\ket{-}\bra{-}\bra{-}$. We can represent a copying process which introduces a phase difference of $\alpha$ between the two basis states by writing $\alpha$ on the spider. In case nothing is written on the spider, it is assumed that the spider introduces no phase difference between the two basis states while copying. The basic elements of $ZX$-calculus are indicated in Fig. \ref{qnoh_Fig7}.\\

\begin{figure}
\centering
\includegraphics[scale=.3]{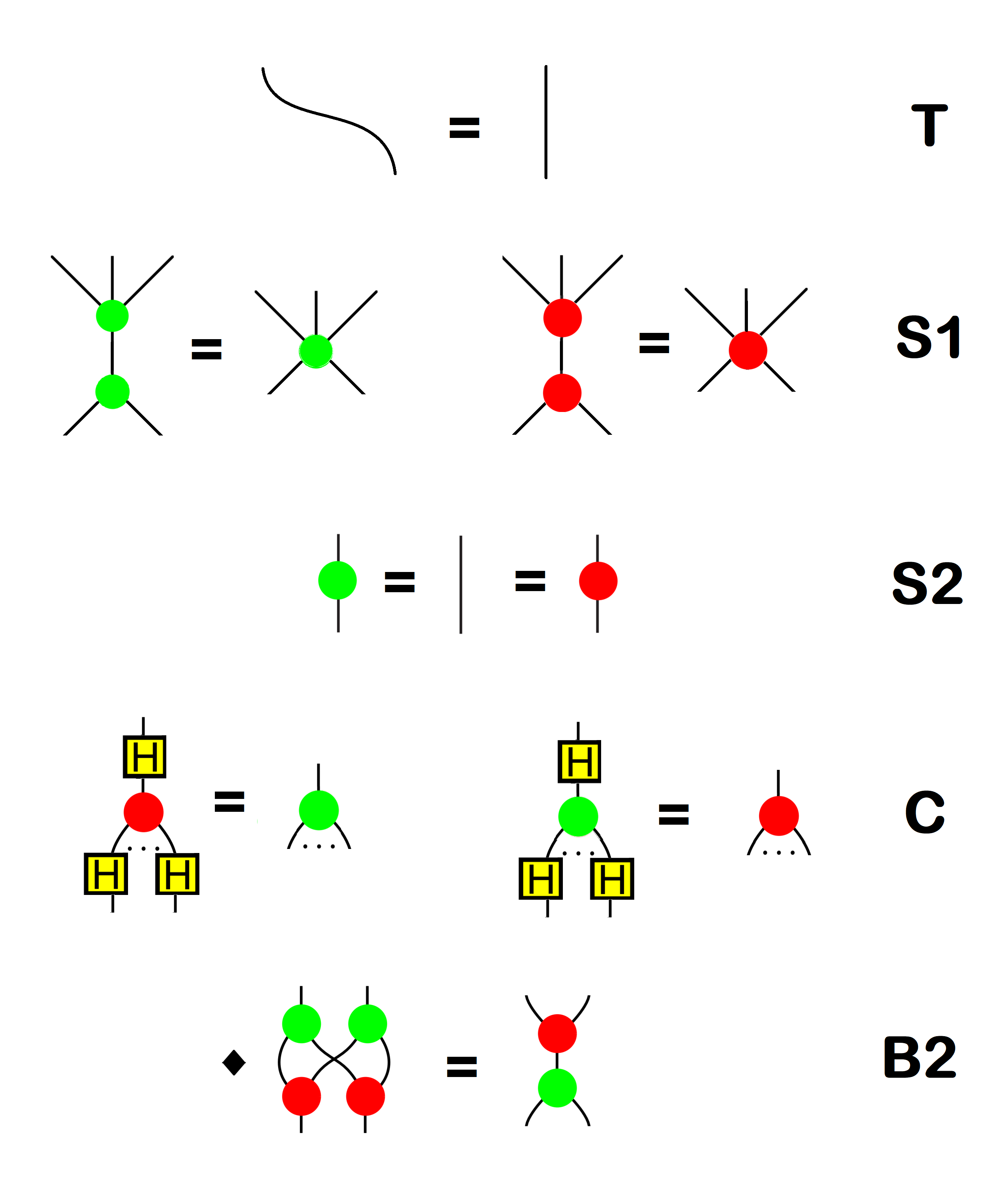}
\caption{Equational rules for the $ZX$-calculus.}
\label{qnoh_Fig8}
\end{figure}

These definitions naturally lead to several simple diagrammatic rules for manipulating circuits consisting only of red and green spiders, and Hadamard gates, summarized in Fig. \ref{qnoh_Fig8}. A more detailed list of rules can be found in Ref. \cite{qnoh_Bob1}. The simplest rule is that changing the ``topology" of wires while maintaining the connections preserves the circuit. Further, it is clear from their definition that spiders of the same colour can be merged together, as in rule $S1$, and that spiders with one input and one output wire are the same as identity, as in rule $S2$. It is also clear from the definition of the Hadamard gate, that it can be used to transform red spiders into green spiders and vice-versa, as in rule $C$. Finally, if we were to express a red dot with two incoming wires and one outgoing wire in the computational basis, one would find it is equivalent to a sum operator $\sum_{a,b}\ket{a\oplus b}\bra{a}\bra{b}$, which naturally leads to rule $B2$, as well as the representation of a CNOT gate shown in Fig. \ref{qnoh_Fig7}. It is also possible to start with the rules depicted in Fig. \ref{qnoh_Fig8} as axioms for a category-theoretic formulation of (a subtheory of) quantum mechanics, but we do not discuss this here.

Now we are ready to begin the diagrammatic derivation. For the ease of illustration, we use a different unitary operator than the one given in Eq. \eqref{qnoh_Eq2}. It should be noted that this performs the same operation as the unitary shown in Eq. \eqref{qnoh_Eq2}.

\begin{equation}\label{qnoh_Eq6}
    \mathbf{U} =  \mathds{1} \otimes \ket{00}\bra{00} + \mathbf{X} \otimes \ket{01}\bra{01} - i \mathbf{Y}\otimes \ket{10}\bra{10}+\mathbf{Z}\otimes \ket{11}\bra{11}
\end{equation}

\begin{figure}
\centering
\includegraphics[scale=.25]{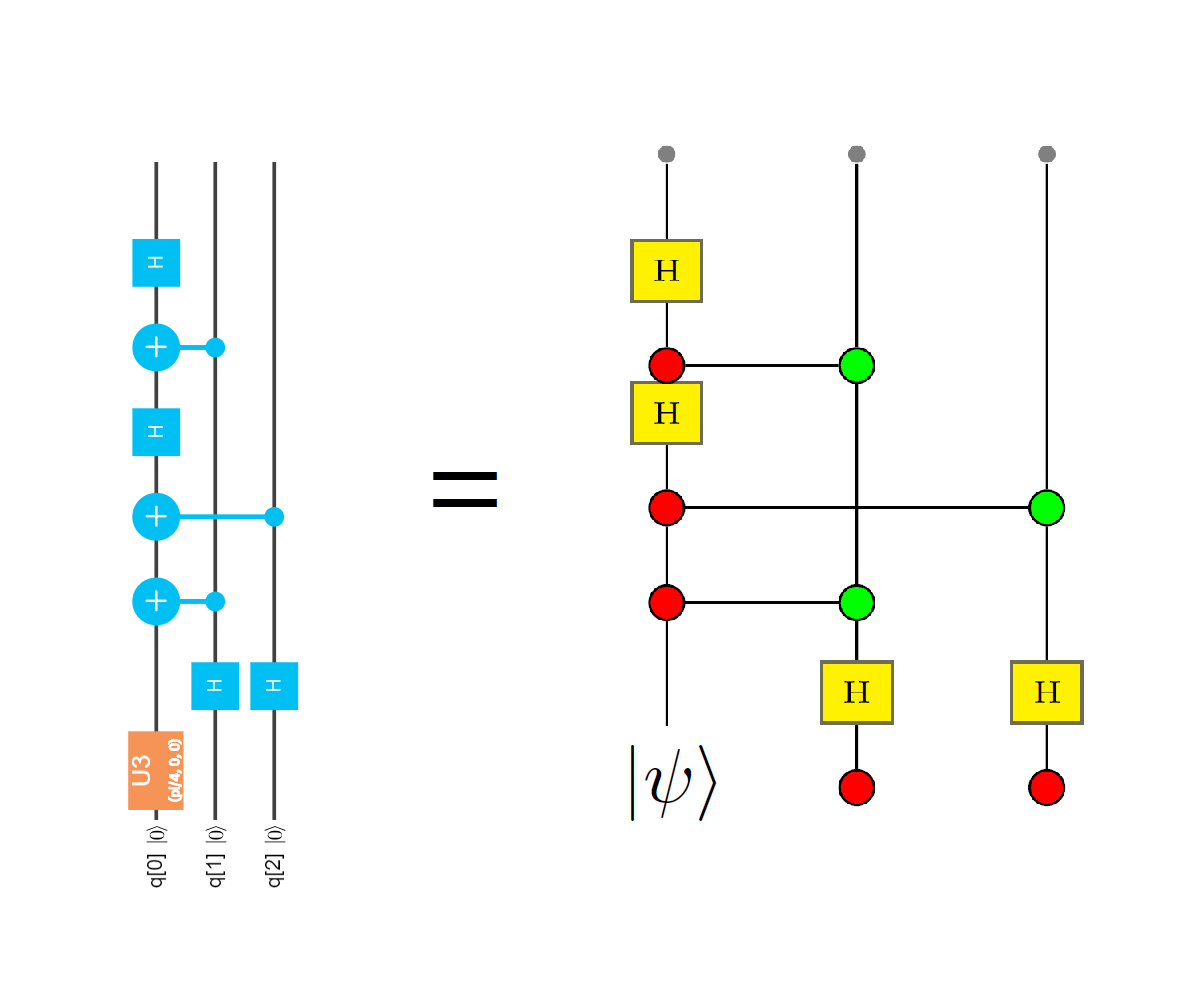}
\caption{Converting the circuit represented by Eq. \eqref{qnoh_Eq3} to the language of $ZX$-calculus}
\label{qnoh_zx}
\end{figure}

It is easy to verify that this unitary performs the desired bleaching. We can express the circuit representing this unitary in the language of $ZX$-calculus as shown in Fig. \ref{qnoh_zx}. By repeated application of the simplification rules shown in Fig. \ref{qnoh_Fig8}, we can trace the flow of quantum information from the original system to the ancillas.

\clearpage

\begin{quote}
\center
\begin{tikzpicture}[baseline={([yshift=-.5ex]current bounding box.center)}]
	\begin{pgfonlayer}{nodelayer}
		\node [style=hadamard] (v4) at (-4.0, -3.0) {};
		\node [style=hadamard] (v16) at (-8.0, 2.5) {};
		\node [style=wire] (b1) at (-6.0, 4.0) {};
		\node [style=wire] (b0) at (-8.0, 4.0) {};
		\node [style=Z] (v12) at (-5.999999, 1.5) {};
		\node [style=hadamard] (v3) at (-6.0, -3.0) {};
		\node [style=Z] (v11) at (-3.999999, 0.0) {};
		\node [style=wire] (b2) at (-4.0, 4.0) {};
		\node [style=X] (v1) at (-5.999999, -4.0) {};
		\node [style=hadamard] (v15) at (-8.0, 0.999) {};
		\node [style=Z] (v10) at (-5.999999, -1.0) {};
		\node [style=wire2] (v0) at (-8.0, -4.0) {$\ket{\psi}$};
		\node [style=X] (v7) at (-7.999999, 1.5) {};
		\node [style=X] (v2) at (-3.999999, -4.0) {};
		\node [style=X] (v5) at (-7.999999, -1.0) {};
		\node [style=X] (v6) at (-7.999999, 0.0) {};
		\node [style=none] (padl) at (-9.0, -5.0) {};
		\node [style=none] (padr) at (-2.999999, 5.0) {};
		\node [style=none] (padu) at (-9.0, 5.0) {};
		\node [style=none] (padd) at (-2.999999, -5.0) {};
	\end{pgfonlayer}
	\begin{pgfonlayer}{edgelayer}
		\draw [style=simple] (v6) to (v15);
		\draw [style=simple] (v11) to (v6);
		\draw [style=simple] (v5) to (v6);
		\draw [style=simple] (v5) to (v10);
		\draw [style=simple] (v2) to (v4);
		\draw [style=simple] (v15) to (v7);
		\draw [style=simple] (v12) to (v7);
		\draw [style=simple] (v7) to (v16);
		\draw [style=simple] (v0) to (v5);
		\draw [style=simple] (v10) to (v12);
		\draw [style=simple] (v1) to (v3);
		\draw [style=simple] (v11) to (b2);
		\draw [style=simple] (v3) to (v10);
		\draw [style=simple] (v16) to (b0);
		\draw [style=simple] (v12) to (b1);
		\draw [style=simple] (v4) to (v11);
	\end{pgfonlayer}
\end{tikzpicture}
$\underset{C}{=}$
\begin{tikzpicture}[baseline={([yshift=-.5ex]current bounding box.center)}]
	\begin{pgfonlayer}{nodelayer}
		\node [style=wire] (b1) at (-6.0, 4.0) {};
		\node [style=wire] (b0) at (-8.0, 4.0) {};
		\node [style=Z] (v12) at (-5.999999, 1.5) {};
		\node [style=Z] (v11) at (-3.999999, 0.0) {};
		\node [style=wire] (b2) at (-4.0, 4.0) {};
		\node [style=Z] (v10) at (-5.999999, -1.0) {};
		\node [style=wire2] (v0) at (-8.0, -4.0) {$\ket{\psi}$};
		\node [style=Z] (v7) at (-3.999999, -4.0) {};
		\node [style=Z] (v13) at (-7.999999, 1.5) {};
		\node [style=Z] (v2) at (-5.999999, -3.999999) {};
		\node [style=X] (v5) at (-7.999999, -1.0) {};
		\node [style=hadamard] (v8) at (-6.999999, 1.5) {};
		\node [style=X] (v6) at (-7.999999, 0.0) {};
		\node [style=none] (padl) at (-9.0, -5.0) {};
		\node [style=none] (padr) at (-2.999999, 5.0) {};
		\node [style=none] (padu) at (-9.0, 5.0) {};
		\node [style=none] (padd) at (-2.999999, -5.0) {};
	\end{pgfonlayer}
	\begin{pgfonlayer}{edgelayer}
		\draw [style=simple] (v6) to (v13);
		\draw [style=simple] (v11) to (v6);
		\draw [style=simple] (v8) to (v13);
		\draw [style=simple] (v12) to (v8);
		\draw [style=simple] (v5) to (v6);
		\draw [style=simple] (v5) to (v10);
		\draw [style=simple] (v2) to (v10);
		\draw [style=simple] (v7) to (v11);
		\draw [style=simple] (v0) to (v5);
		\draw [style=simple] (v10) to (v12);
		\draw [style=simple] (v11) to (b2);
		\draw [style=simple] (v13) to (b0);
		\draw [style=simple] (v12) to (b1);
	\end{pgfonlayer}
\end{tikzpicture}
$\underset{S1}{=}$
\begin{tikzpicture}[baseline={([yshift=-.5ex]current bounding box.center)}]
	\begin{pgfonlayer}{nodelayer}
		\node [style=Z] (v9) at (-4.0, 0.0) {};
		\node [style=wire] (b1) at (-6.0, 4.0) {};
		\node [style=wire] (b0) at (-8.0, 4.0) {};
		\node [style=Z] (v12) at (-5.999999, 1.5) {};
		\node [style=Z] (v3) at (-6.0, -1.0) {};
		\node [style=wire] (b2) at (-4.0, 4.0) {};
		\node [style=wire2] (v0) at (-8.0, -4.0) {$\ket{\psi}$};
		\node [style=Z] (v13) at (-7.999999, 1.5) {};
		\node [style=X] (v5) at (-7.999999, -1.0) {};
		\node [style=hadamard] (v8) at (-6.999999, 1.5) {};
		\node [style=X] (v6) at (-7.999999, 0.0) {};
		\node [style=none] (padl) at (-9.0, -5.0) {};
		\node [style=none] (padr) at (-3.0, 5.0) {};
		\node [style=none] (padu) at (-9.0, 5.0) {};
		\node [style=none] (padd) at (-3.0, -5.0) {};
	\end{pgfonlayer}
	\begin{pgfonlayer}{edgelayer}
		\draw [style=simple] (v6) to (v13);
		\draw [style=simple] (v9) to (v6);
		\draw [style=simple] (v8) to (v13);
		\draw [style=simple] (v12) to (v8);
		\draw [style=simple] (v5) to (v6);
		\draw [style=simple] (v0) to (v5);
		\draw [style=simple] (v9) to (b2);
		\draw [style=simple] (v5) to (v3);
		\draw [style=simple] (v3) to (v12);
		\draw [style=simple] (v13) to (b0);
		\draw [style=simple] (v12) to (b1);
	\end{pgfonlayer}
\end{tikzpicture}
$\underset{S1}{=}$
\begin{tikzpicture}[baseline={([yshift=-.5ex]current bounding box.center)}]
	\begin{pgfonlayer}{nodelayer}
		\node [style=Z] (v9) at (-4.0, 0.0) {};
		\node [style=wire] (b1) at (-6.0, 4.0) {};
		\node [style=wire] (b0) at (-8.0, 4.0) {};
		\node [style=Z] (v12) at (-5.999999, 1.5) {};
		\node [style=Z] (v3) at (-6.0, -1.0) {};
		\node [style=wire] (b2) at (-4.0, 4.0) {};
		\node [style=X] (v1) at (-7.999999, -1.0) {};
		\node [style=wire2] (v0) at (-8.0, -4.0) {$\ket{\psi}$};
		\node [style=Z] (v13) at (-7.999999, 1.5) {};
		\node [style=hadamard] (v8) at (-6.999999, 1.5) {};
		\node [style=none] (padl) at (-9.0, -5.0) {};
		\node [style=none] (padr) at (-3.0, 5.0) {};
		\node [style=none] (padu) at (-9.0, 5.0) {};
		\node [style=none] (padd) at (-3.0, -5.0) {};
	\end{pgfonlayer}
	\begin{pgfonlayer}{edgelayer}
		\draw [style=simple] (v8) to (v13);
		\draw [style=simple] (v12) to (v8);
		\draw [style=simple] (v1) to (v0);
		\draw [style=simple] (v1) to (v13);
		\draw [style=simple] (v1) to (v3);
		\draw [style=simple] (v9) to (v1);
		\draw [style=simple] (v9) to (b2);
		\draw [style=simple] (v3) to (v12);
		\draw [style=simple] (v13) to (b0);
		\draw [style=simple] (v12) to (b1);
	\end{pgfonlayer}
\end{tikzpicture}
$\underset{S1}{=}$
\begin{tikzpicture}[baseline={([yshift=-.5ex]current bounding box.center)}]
	\begin{pgfonlayer}{nodelayer}
		\node [style=Z] (v9) at (-4.0, 0.0) {};
		\node [style=wire] (b1) at (-6.0, 4.0) {};
		\node [style=wire] (b0) at (-8.0, 4.0) {};
		\node [style=wire] (b2) at (-4.0, 4.0) {};
		\node [style=X] (v1) at (-7.999999, -1.0) {};
		\node [style=wire2] (v0) at (-8.0, -4.0) {$\ket{\psi}$};
		\node [style=Z] (v13) at (-7.999999, 1.5) {};
		\node [style=Z] (v2) at (-6.0, 1.5) {};
		\node [style=hadamard] (v8) at (-6.999999, 1.5) {};
		\node [style=none] (padl) at (-9.0, -5.0) {};
		\node [style=none] (padr) at (-3.0, 5.0) {};
		\node [style=none] (padu) at (-9.0, 5.0) {};
		\node [style=none] (padd) at (-3.0, -5.0) {};
	\end{pgfonlayer}
	\begin{pgfonlayer}{edgelayer}
		\draw [style=simple] (v8) to (v13);
		\draw [style=simple] (v2) to (v8);
		\draw [style=simple] (v1) to (v0);
		\draw [style=simple] (v1) to (v2);
		\draw [style=simple] (v1) to (v13);
		\draw [style=simple] (v9) to (v1);
		\draw [style=simple] (v9) to (b2);
		\draw [style=simple] (v13) to (b0);
		\draw [style=simple] (b1) to (v2);
	\end{pgfonlayer}
\end{tikzpicture}
$\underset{T}{=}$
\begin{tikzpicture}[baseline={([yshift=-.5ex]current bounding box.center)}]
	\begin{pgfonlayer}{nodelayer}
		\node [style=wire] (b1) at (-6.0, 4.0) {};
		\node [style=wire] (b0) at (-8.0, 4.0) {};
		\node [style=wire] (b2) at (-4.0, 4.0) {};
		\node [style=X] (v1) at (-5.999999, -1.3333333333333333) {};
		\node [style=wire2] (v0) at (-8.0, -4.0) {$\ket{\psi}$};
		\node [style=Z] (v13) at (-7.999999, 1.5) {};
		\node [style=Z] (v2) at (-6.0, 1.5) {};
		\node [style=hadamard] (v8) at (-6.999999, 1.5) {};
		\node [style=none] (padl) at (-9.0, -5.0) {};
		\node [style=none] (padr) at (-3.0, 5.0) {};
		\node [style=none] (padu) at (-9.0, 5.0) {};
		\node [style=none] (padd) at (-3.0, -5.0) {};
	\end{pgfonlayer}
	\begin{pgfonlayer}{edgelayer}
		\draw [style=simple] (v8) to (v13);
		\draw [style=simple] (v2) to (v8);
		\draw [style=simple] (v1) to (v0);
		\draw [style=simple] (v1) to (v2);
		\draw [style=simple] (v1) to (v13);
		\draw [style=simple] (b2) to (v1);
		\draw [style=simple] (v13) to (b0);
		\draw [style=simple] (b1) to (v2);
	\end{pgfonlayer}
\end{tikzpicture}
$\underset{T,S1}{=}$
\begin{tikzpicture}[baseline={([yshift=-.5ex]current bounding box.center)}]
	\begin{pgfonlayer}{nodelayer}
		\node [style=Z] (v4) at (-8.0, -1.999) {};
		\node [style=wire] (b1) at (-6.0, 4.0) {};
		\node [style=wire] (b0) at (-8.0, 4.0) {};
		\node [style=Z] (v3) at (-6.0, 0.0) {};
		\node [style=wire] (b2) at (-4.0, 4.0) {};
		\node [style=X] (v1) at (-3.999999, 1.6666666666666667) {};
		\node [style=wire2] (v0) at (-8.0, -4.0) {$\ket{\psi}$};
		\node [style=Z] (v2) at (-6.0, 2.999) {};
		\node [style=hadamard] (v8) at (-5.999999, 1.5) {};
		\node [style=none] (padl) at (-9.0, -5.0) {};
		\node [style=none] (padr) at (-2.999999, 5.0) {};
		\node [style=none] (padu) at (-9.0, 5.0) {};
		\node [style=none] (padd) at (-2.999999, -5.0) {};
	\end{pgfonlayer}
	\begin{pgfonlayer}{edgelayer}
		\draw [style=simple] (v2) to (v8);
		\draw [style=simple] (v1) to (v0);
		\draw [style=simple] (v1) to (v2);
		\draw [style=simple] (v1) to (v3);
		\draw [style=simple] (b2) to (v1);
		\draw [style=simple] (v8) to (v3);
		\draw [style=simple] (v3) to (v4);
		\draw [style=simple] (v4) to (b0);
		\draw [style=simple] (b1) to (v2);
	\end{pgfonlayer}
\end{tikzpicture}
$\underset{S1}{=}$
\begin{tikzpicture}[baseline={([yshift=-.5ex]current bounding box.center)}]
	\begin{pgfonlayer}{nodelayer}
		\node [style=Z] (v4) at (-8.0, -1.999) {};
		\node [style=wire] (b1) at (-6.0, 4.0) {};
		\node [style=wire] (b0) at (-8.0, 4.0) {};
		\node [style=Z] (v3) at (-6.0, 0.0) {};
		\node [style=wire] (b2) at (-4.0, 4.0) {};
		\node [style=wire2] (v0) at (-8.0, -4.0) {$\ket{\psi}$};
		\node [style=Z] (v2) at (-6.0, 2.999) {};
		\node [style=X] (v5) at (-3.999, 2.999) {z};
		\node [style=hadamard] (v8) at (-5.999999, 1.5) {};
		\node [style=X] (v6) at (-3.999, 0.0) {-z};
		\node [style=none] (padl) at (-9.0, -5.0) {};
		\node [style=none] (padr) at (-2.999, 5.0) {};
		\node [style=none] (padu) at (-9.0, 5.0) {};
		\node [style=none] (padd) at (-2.999, -5.0) {};
	\end{pgfonlayer}
		\draw [style=simple] (v5) to (v6);	\begin{pgfonlayer}{edgelayer}

		\draw [style=simple] (v2) to (v8);
		\draw [style=simple] (v5) to (v2);
		\draw [style=simple] (v6) to (v0);
		\draw [style=simple] (b2) to (v5);
		\draw [style=simple] (v6) to (v3);
		\draw [style=simple] (v8) to (v3);
		\draw [style=simple] (v3) to (v4);
		\draw [style=simple] (v4) to (b0);
		\draw [style=simple] (b1) to (v2);
	\end{pgfonlayer}
\end{tikzpicture}
\end{quote}

As can be clearly seen from the final figure, the state $\ket{\psi}$ is encoded in the two ancilla qubits, while the first qubit now contains no information about $\ket{\psi}$. From the analysis, we also see that by applying the CNOT(2,3), Hadamard(2), CNOT(2,3) gates, we can decode the state $\ket{\psi}$ from the two ancilla qubits. The above figures were created using the software \textit{Quantomatic} \cite{qnoh_Kissinger2015}.

While the above derivation demonstrates the flow of information, it would be very interesting to provide a purely diagrammatic (or category-theoretic) proof of the no-hiding theorem, perhaps similar in spirit to the category theoretic versions of the no-cloning theorem \cite{qnoh_abramsky2010no}. It would also be very educational to understand the no-hiding theorem in the context of the systems-theoretic framework of \cite{qnoh_srivastava2011graph}, where quantum information acts as a ``through" variable, analogous to current in an electrical circuit, and obeys an analogue of Kirchoff's current law.

\section{Conclusion}
In this paper, we have demonstrated the experimental verification of the no-hiding theorem using IBM's 5-qubit quantum computer. The ZX calculus has been utilized to obtain the decoding circuit required for illustrating this theorem. Quantum state tomography has been performed to check the accuracy of the implementation. A number of directions can be pursued in future. One of them is the extension of the no-hiding theorem to imperfect hiding processes. We can also investigate the applications of this theorem for retrieving information from a noisy environment.    

\section*{Acknowledgements}
We are extremely grateful to the IBM team and the IBM QE project. The discussions in this paper do not reflect any opinion of the IBM QE team or IBM. ARK thanks IISER Kolkata for hospitality during which part of this work was completed. ARK and NG thank the National Initiative for Undergraduate Science (NIUS) Physics for support. ARK also acknowledges an AADEIs Undergraduate Research Award (UGRA) which partially supported this work in its final stages. BKB acknowledges Institute Fellowship provided by IISER-K. SP thanks ICTS, Bengaluru for hospitality, and acknowledges the support of a DST-SERB Early Career Research
Award (ECR/2017/001023) and a DST INSPIRE Faculty Award.

\makeatletter
\providecommand \@ifxundefined [1]{%
 \@ifx{#1\undefined}
}%
\providecommand \@ifnum [1]{%
 \ifnum #1\expandafter \@firstoftwo
 \else \expandafter \@secondoftwo
 \fi
}%
\providecommand \@ifx [1]{%
 \ifx #1\expandafter \@firstoftwo
 \else \expandafter \@secondoftwo
 \fi
}%
\providecommand \natexlab [1]{#1}%
\providecommand \enquote  [1]{``#1''}%
\providecommand \bibnamefont  [1]{#1}%
\providecommand \bibfnamefont [1]{#1}%
\providecommand \citenamefont [1]{#1}%
\providecommand \href@noop [0]{\@secondoftwo}%
\providecommand \href [0]{\begingroup \@sanitize@url \@href}%
\providecommand \@href[1]{\@@startlink{#1}\@@href}%
\providecommand \@@href[1]{\endgroup#1\@@endlink}%
\providecommand \@sanitize@url [0]{\catcode `\\12\catcode `\$12\catcode
  `\&12\catcode `\#12\catcode `\^12\catcode `\_12\catcode `\%12\relax}%
\providecommand \@@startlink[1]{}%
\providecommand \@@endlink[0]{}%
\providecommand \url  [0]{\begingroup\@sanitize@url \@url }%
\providecommand \@url [1]{\endgroup\@href {#1}{\urlprefix }}%
\providecommand \urlprefix  [0]{URL }%
\providecommand \Eprint [0]{\href }%
\providecommand \doibase [0]{http://dx.doi.org/}%
\providecommand \selectlanguage [0]{\@gobble}%
\providecommand \bibinfo  [0]{\@secondoftwo}%
\providecommand \bibfield  [0]{\@secondoftwo}%
\providecommand \translation [1]{[#1]}%
\providecommand \BibitemOpen [0]{}%
\providecommand \bibitemStop [0]{}%
\providecommand \bibitemNoStop [0]{.\EOS\space}%
\providecommand \EOS [0]{\spacefactor3000\relax}%
\providecommand \BibitemShut  [1]{\csname bibitem#1\endcsname}%
\let\auto@bib@innerbib\@empty


\begin{thebibliography}{66}
\bibitem{qnoh_IBM} IBM Q, URL \url{http://research.ibm.com/ibm-q/}.
\bibitem{qnoh_LeggetIBM} Huffman E., Mizel A.:Violation of noninvasive macrorealism by a superconducting qubit: Implementation of a Leggett-Garg test that addresses the clumsiness loophole. Phys. Rev. A
\bibitem{qnoh_MerminIBM} Alsina, D., Latorre, J. L.: Experimental test of Mermin inequalities on a five-qubit quantum computer. Phys. Rev. A
\textbf{94}, 012314 (2016)
\bibitem{qnoh_cond1} Dumitrescu E.F,, McCaskey A.J,, Hagen G., Jansen G.R., Morris T.D,, Papenbrock T., Pooser R.C,, Dean D.J., Lougovski P.: Cloud Quantum Computing of an Atomic Nucleus. Phys. Rev. Lett. \textbf{120},210501 (2018)
\bibitem{qnoh_cond2} Choo K., von Keyserlingk C.W., Regnault N., Neupert.T: Measurement of the entanglement spectrum of a symmetry-protected topological state using the IBM quantum computer.
Phys. Rev. Lett. \textbf{121}, 086808 (2018)
\bibitem{qnoh_ai1} Zhao Z., Kerstjens A.P., Rebentrost P. Wittek P.: Bayesian Deep Learning on a Quantum Computer, 	arXiv:1806.11463
\bibitem{qnoh_qg1} Manabputra, Behera, B.~K., Panigrahi, P.~K.: A Simulational Model for Witnessing Quantum Effects of Gravity Using IBM Quantum Computer. arXiv:1806.10229 (2018)
\bibitem{npjQIViyuela2018} Viyuela, O., Rivas, A., Gasparinetti, S., Wallraff, A., Filipp, S., Martin-Delgado, M.~A.: Observation of topological Uhlmann phases with superconducting qubits. npj Quantum Inf. \textbf{4}, 10 (2018) 
\bibitem{qnoh_qg2} Kapil M., Behera, B.~K., Panigrahi, P.~K.: Quantum Simulation of Klein Gordon Equation and Observation of Klein Paradox in IBM Quantum Computer. arXiv:1807.00521 (2018)
\bibitem{qnoh_BeheraQIP2017} Behera, B.~K., Banerjee, A., Panigrahi, P.~K.: Experimental realization of quantum cheque using a five-qubit quantum computer. Quantum Inf. Process. \textbf{16}, 312 (2017) 
\bibitem{qnoh_RGBiswas2018} Biswas, S., Razdan, S., Behera, B.~K., Panigrahi, P.~K.: Realization of Counterfactual Quantum Cryptography Us-ing IBM's Quantum Computer. DOI: 10.13140/RG.2.2.30090.52160 (2018) 
\bibitem{qnoh_WoottonIBM} Wootton, J.R: Demonstrating non-Abelian braiding of surface code defects in a five qubit experiment. Quantum Sci. Technol. \textbf{2}, 015006 (2017)
\bibitem{qnoh_Bikash3} Ghosh D., Agarwal P., Pandey P., Behera, B.~K., Panigrahi, P.~K.: Automated error correction in IBM quantum computer and explicit generalization. . Quantum Inf Process \textbf{17}, 153 (2018)
\bibitem{qnoh_VuillotQIC2018} Vuillot, C.: Is error detection helpful on IBM 5Q chips ? Quantum Inf. Comput. \textbf{18}, 0949-0964 (2018) 
\bibitem{qnoh_arXivHarper2018} Harper, R., Flammia, S.: Fault tolerance in the IBM Q Experience. arXiv:1806.02359 (2018)
\bibitem{qnoh_SisodiaQIP2017} Sisodia, M., Verma, V., Thapliyal, K., Pathak, A.: Teleportation of a qubit using entangled non-orthogonal states: a comparative study. Quantum Inf. Process. \textbf{16}, 76 (2017)
\bibitem{qnoh_Anirban} Sisodia M., Shukla A., Pathak A.: Experimental realization of nondestructive discrimination of Bell states using a five-qubit quantum computer. Phys. Lett. A \textbf{381}, 3860 (2017)
\bibitem{qnoh_Abhishek} Sisodia M., Shukla A., Thapliyal K., Pathak A.: Design and experimental realization of an optimal scheme for teleportation of an n-qubit quantum state. Quantum Inf. Process. \textbf{16}, 292 (2017) 
\bibitem{qnoh_EntropicIBM} Berta M., Wehner S., Wilde, M. M.: Entropic uncertainty and measurement reversibility. New J. Phys. \textbf{18} 073004 (2016)
Proceedings of the National Academy of Sciences Mar 2017, 114 (13) 3305-3310

\bibitem{qnoh_Bikash2} Swain M., Rai A.,  Behera, B.~K., Panigrahi, P.~K.: Experimental demonstration of the violations of Mermin's and Svetlichny's inequalities for W- and GHZ-class of states. arXiv:1810.00874

\bibitem{qnoh_CloudIBM} Devitt J.D.:Performing quantum computing experiments in the cloud
 Phys. Rev. A \textbf{94}, 032329 (2016)
\bibitem{qnoh_LinkeIBM} Figgatt, C., Landsmam, K. A., Wright, K., Monroe, C.: Experimental comparison of two quantum computers.

\bibitem{qnoh_no-cloning}Wootters, W.K., Zurek, W.H.: A single quantum cannot be cloned. Nature\textbf{299}, 802 (1982)
 
\bibitem{qnoh_no-deleting} Pati, A.K., Braunstein, S.L: Impossibility of deleting an unknown quantum state. Nature \textbf{404}, 164-165 (2000)

\bibitem{qnoh_no-hiding} Braunstein, S.L, Pati, A.K: Quantum Information Cannot Be Completely Hidden in Correlations: Implications for the Black-Hole Information Paradox.
Phys. Rev. Lett. \textbf{98}, 080502. (2016)
\bibitem{qnoh_thermalization}  Popescu, S., Short, A.J, Winter, A.: Entanglement and the foundations of statistical mechanics. Nat. Phys. Volume \textbf{2} pages 754–758 (2006)
\bibitem{qnoh_HwakingNature1974} Hawking, S.~W.: Black hole explosions? Nature \textbf{248}, 30 (1974)
\bibitem{qnoh_private} Ambainis, A., Mosca, M., Tapp, A., Wolf R.D.: Private Quantum Channels. Proceedings of the 41st Annual Symposium on Foundations of Computer Science. (2000)

\bibitem{qnoh_no-hiding-NMR} Samal, J.R, Pati, A.K,, Kumar, A.: Experimental Test of the Quantum No-Hiding Theorem. Phys. Rev. Lett. \textbf{106}, 080401.
\bibitem{qnoh_one-time-pad} Vernam, G. S.: Cipher Printing Telegraph Systems For Secret Wire and Radio Telegraphic Communications.Transactions of the American Institute of Electrical Engineers, vol. XLV, (1926)

\bibitem{qnoh_shannon1949} Shannon, C. E.: Communication theory of secrecy systems.
The Bell System Technical Journal. (1949)

\bibitem{qnoh_Bob1} Coecke, B., Duncan, R.: Interacting quantum observables: categorical algebra and diagrammatics. New J. Phys. \textbf{13}, 043016 (2011)

\bibitem{qnoh_Kissinger2015} Kissinger, A., Zamdzhiev, V.Quantomatic: A Proof Assistant for Diagrammatic Reasoning. In: Felty A., Middeldorp A. (eds) Automated Deduction - CADE-25. CADE 2015. Lecture Notes in Computer Science, vol 9195. Springer, Cham (2015)


\bibitem{qnoh_abramsky2010no} Samson, A.: No-Cloning In Categorical Quantum Mechanics. arXiv:0910.2401 [quant-ph]

\bibitem{qnoh_srivastava2011graph}Srivastava, D.P., Sahni, V., Satsangi, P.S.: Graph-theoretic quantum system modelling for information/computation processing circuits, International Journal of General Systems.40:8, 777-804 (2011).

\end{thebibliography}
\end{document}